\documentclass[12pt]{iopart}
\usepackage{graphicx}%
\usepackage{multirow}%
\usepackage{multicol}%
\usepackage{color}%
\usepackage{booktabs}%
\usepackage{subcaption}%
\usepackage[title]{appendix}%
\usepackage{xcolor}%
\usepackage{textcomp}%
\usepackage{manyfoot}%
\usepackage{algorithm}%
\usepackage{algorithmicx}%
\usepackage{algpseudocode}%
\usepackage{listings}%
\usepackage{epstopdf}
\usepackage{bm}
\usepackage{setspace}
\usepackage{here}
\usepackage[colorlinks,linkcolor=blue,citecolor=blue]{hyperref}
\usepackage{amssymb}

\begin{document}

\title[Unusual spin dynamics in the van der Waals antiferromagnet FeGa$_2$S$_4$]{Unusual spin dynamics in the van der Waals antiferromagnet FeGa$_2$S$_4$}
	
\author{Yifei Tang$^1$, Yoshihiko Umemoto$^2$, Yo Kawamoto$^1$, Masahiro Kawamata$^1$, Shinichiro Asai$^3$, Yoichi Ikeda$^2$, Masaki Fujita$^2$, and Yusuke Nambu$^{2,4,5,\ast}$}
\address{$^1$ Department of Physics, Tohoku University, Sendai 980-8578, Japan}
\address{$^2$ Institute for Materials Research, Tohoku University, Sendai 980-8577, Japan}
\address{$^3$ Institute for Solid State Physics, University of Tokyo, Kashiwa, Chiba 277-8581, Japan}
\address{$^4$ Organization for Advanced Studies, Tohoku University, Sendai 980-8577, Japan}
\address{$^5$ FOREST, Japan Science and Technology Agency, Saitama 332-0012, Japan}
\ead{$^{\ast}$nambu@tohoku.ac.jp}

\begin{abstract}
	Spin dynamics in the van der Waals antiferromagnet FeGa$_2$S$_4$ with triangular lattices are investigated using magnetometry, neutron scattering, and muon spin relaxation measurements.
	The characteristic spin relaxation time is thoroughly clarified over thirteen orders of magnitude.
	Although the temperature dependence of DC and AC susceptibilities recalls a conventional spin-glass transition, nonlinear susceptibilities showing no divergences at the anomalous temperature, $T^{\ast}=16.87(7)$~K, deny that and instead hint at other mechanisms.
	Elastic neutron scattering together with previously measured muon results depict a slowly fluctuated ($\sim 10^{-5}$~sec) spin state above $T^{\ast}$.
	In juxtaposing the underlying simplest structure among frustrated magnets with an intricate hierarchy of time scales, FeGa$_2$S$_4$ can be a playground for studying temporal spin correlations in the two-dimensional limit.
\end{abstract}		

\vspace{2pc}

%\submitto{\JPCM}
\maketitle

\ioptwocol

%\section{Introduction}

Unconventional magnetism has always been the major issue in condensed matter physics, for which spatial and spin dimensionality have considerable influences~\cite{Wang2022ACS,Ningrum2020Res}.
In recent decades, particular interest has been caught in the two-dimensional (2D) systems~\cite{Burch2018Nat, Jiang2021APR}.
Whereas the theorem~\cite{Mermin1966PRL} prohibits the magnetic long-range order (LRO) in $d\le 2$ dimensional materials with continuous spins, unusual spin fluctuations and quantum effects have been realised in 2D triangular lattice antiferromagnets~\cite{CapriottiPRL1999}.
Triangular antiferromagnets, having the simplest structure among geometrically frustrated magnets~\cite{Ramirez2001Book,DiepWC}, can foster a noncollinear arrangement of spins.
The noncollinearity can naturally couple with the chiral degree of freedom~\cite{Kawamura1984JPSJ,Okubo2010JPSJ}, which gives rise to unconventional magnetism~\cite{Aoyama2020PRL}.
The ground state is yet believed to be in the 120$^{\circ}$ structure~\cite{Huse1988PRL,BernuPRL1992}, but longer-range interactions and other ingredients can induce a wide variety of phases~\cite{kanoda2011mott}.
Among inorganic materials, two model compounds, NiGa$_2$S$_4$~\cite{Nakatsuji2005Sci,Nambu2015PRL,Nambu2009PRB,Nambu2008PRL} and its sibling FeGa$_2$S$_4$~\cite{Nakatsuji2007PRL}, have garnered significant interest given that the ideal two-dimensionality is realised owing to the van der Waals gap.

\begin{figure}[t]
	\centering
	\includegraphics[width=\linewidth,bb=0 0 788 537]{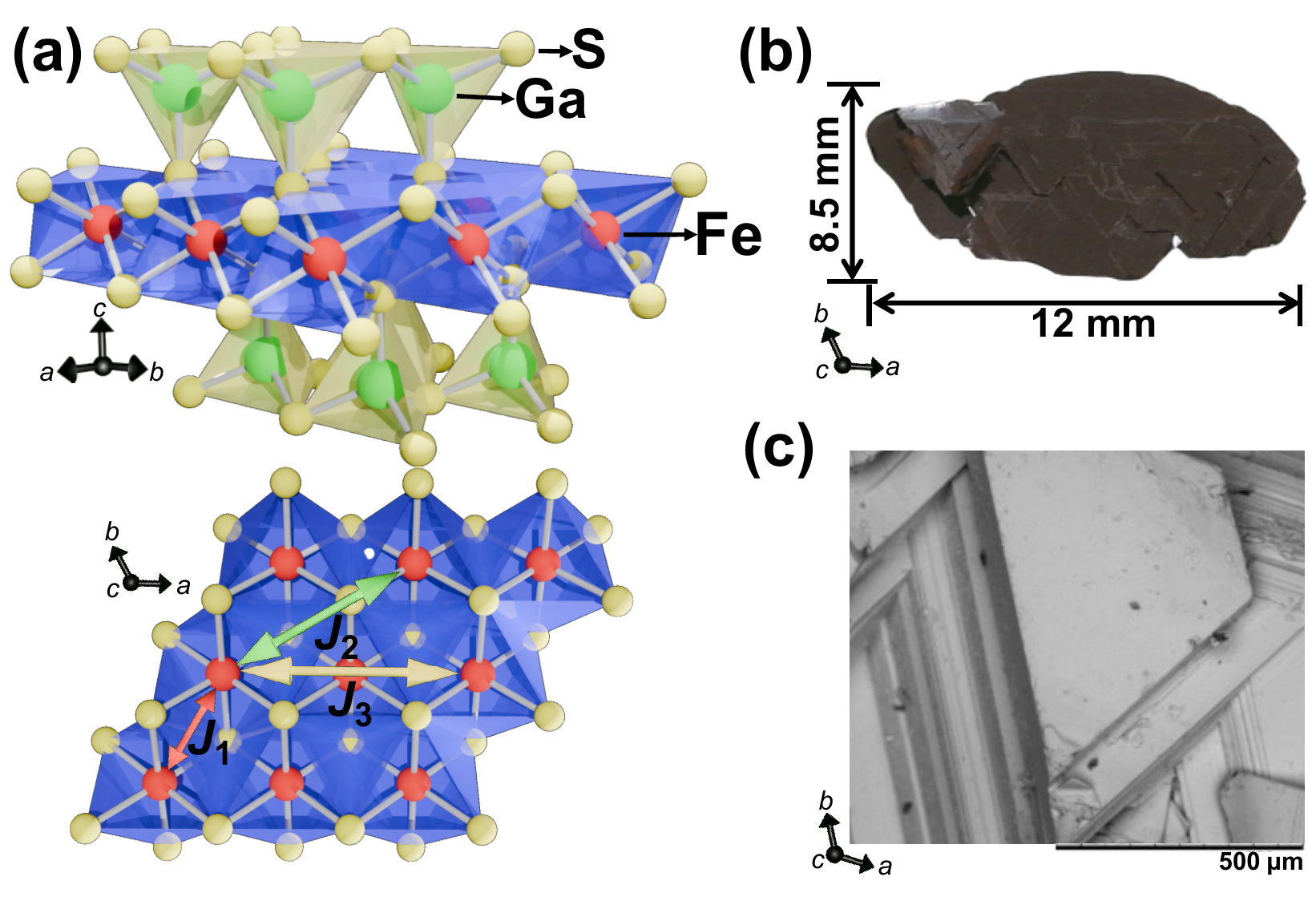}
	\caption{(a) Crystal structures of FeGa$_2$S$_4$ with arrows marking the atomic sites (top) and superexchange interaction pathways (see main text) within a triangular lattice plane (bottom). (b) Photo of a typical piece of FeGa$_2$S$_4$ single crystal. (c) Scanning electron microscopy picture obtained on the surface of FeGa$_2$S$_4$ single crystal.}
	\label{Sam}
\end{figure}

$A$Ga$_2$S$_4$ ($A$ = Ni, Fe) crystalises into the space group $P\bar{3}m1$ [Fig.~\ref{Sam}(a)].
The magnetic atom ($A$) forms an equilateral triangular lattice on the $ab$-plane, where the van der Waals force curtails electrons hopping beyond layers, leading to sufficient two-dimensionality.
$A$S$_6$ octahedral coordinate activates the ligand field, where Ni$^{2+}$ ions have $e_{\rm g}^2$$t_{\rm 2g}^6$ electronic configuration with $S=1$~\cite{Takubo2007PRL}, and Fe$^{2+}$ ions $e_{\rm g}^2$$t_{\rm 2g}^4$ with $S=2$~\cite{Takubo2009PRB}.
Both compounds behave semiconducting with the electronic charge gap of 0.2--0.3~eV~\cite{Tomita2009JPSJ}. 

The materials $A$Ga$_2$S$_4$ do not exhibit conventional magnetic LRO down to the lowest measured temperature.
In NiGa$_2$S$_4$, as separated from the gradual growth of the spatial correlation with decreasing temperature, temporal spin correlation indicates peculiar dynamics below $T^{\ast}=8.5$~K, which is defined by the cusp of DC suscepibility~\cite{Nambu2015PRL}. 
Frequency-dependent peaks of AC susceptibility at around 4~K gives the lower bound, where megahertz spin fluctuations exist in the regime for $4~\rm{K}<$ $T\leq T^{\ast}$~\cite{Nambu2015PRL}.
On the other hand, $T^{\ast}\sim16$ K is found in FeGa$_2$S$_4$~\cite{Nakatsuji2007PRL}.
Both compounds have the scaled temperatures, $T^{\ast}\sim|\theta_{\rm W}|/10$ with $\theta_{\rm W}$ being the Weiss temperature, indicating strong in-plane frustration~\cite{Nakatsuji2007PRL}.

Whereas the observed bifurcations in the DC susceptibility below $T^{\ast}$ and the frequency-dependence of the AC susceptibility are typical for conventional spin-glass (SG) systems, the $T^2$-dependent magnetic specific heat ($C_{\rm M}$) are observed in both compounds~\cite{Nakatsuji2007PRL}, which contrasts with the linear-$T$ dependence in conventional SG systems. 
The discrepancy implies that an additional mechanism is needed to explain the $T^{\ast}$ anomaly.
Low-temperature magnetic state appears to be preserved only for the integer size of spins including FeGa$_2$S$_4$ with $S=2$~\cite{Nambu2008PRL,Nambu2006JPSJ}, where other diluted systems with half-integer spins violate it~\cite{Nambu2011}. 
The spin-size-dependent magnetism points to macroscopic quantum effects inherent, typically modelled via the biquadratic term in the spin Hamiltonian. 
Recently, such biquadratic correlations have been detected using Raman scattering~\cite{Valentine2020}.

Despite $A$Ga$_2$S$_4$ share the above-mentioned low-temperature magnetism, they behave differently in some aspects.
The muon spin relaxation ($\mu$SR) study on FeGa$_2$S$_4$~\cite{Zhao2012PRB} reveals a critical slowing down of spins towards $T_{\mu}=31(2)$~K being nearly twice higher than $T^{\ast}$, whereas it directly goes towards $T^{\ast}$ in the Ni case ($T_{\mu}=9.0(5)$~K $\sim T^{\ast}$).
Unlike spin dynamics realised in $A$Ga$_2$S$_4$, which are supposed to arise from differences in the spin size and the orbital degree of freedom, it is crucial to further elucidate the detailed temporal correlations.
We here argue that slow spin fluctuations are also observable in FeGa$_2$S$_4$ via magnetometry, neutron, and muon results using single crystalline samples.

%\section{Methods}

Polycrystalline samples of FeGa$_2$S$_4$ were synthesised by the solid-state reaction using raw materials with the molar ratio of ${\rm Fe}:{\rm Ga}:{\rm S} = 1:2:4.04$~\cite{Nambu2008JCG}.
We put 1\% sulphur additionally to compensate for the possible loss during the synthesis~\cite{Nambu2009PRB}.
The raw materials were sealed into a quartz ampule, reacted at 1000$^{\circ}$C, and then homogenised by grinding.
The ground samples were resealed and reacted again at 1000$^{\circ}$C.

The obtained polycrystalline samples were put into ampules to grow single crystals by the chemical vapour transport method~\cite{Pardo1981MRB}.
The transport agent was iodine with a concentration of 3~mg/cm$^3$, and the temperature gradient between 925 and 1000$^{\circ}$C was selected.
The grown single crystals are thin and plate-like shapes with a metallic lustre and size up to $\sim$1~cm [Fig.~\ref{Sam}(b)].
Naturally cut triangle edges on the sample surface reflect the symmetry of the space group [Fig.~\ref{Sam}(c)].
The obtained single crystals were purified by annealing under a sulphur atmosphere at 600$^{\circ}$C to compensate for possible sulphur deficiency.

The quality of the single crystals was confirmed by x-ray diffraction and scanning electron microscopy with energy-dispersive X-ray analysis (SEM-EDX) measurements.
DC, AC, and nonlinear susceptibilities were measured using a superconducting quantum interference device magnetometer in the Magnetic Property Measurement System manufactured by Quantum Design Inc.

Elastic channel data were taken on the neutron triple-axis spectrometers, C1-1 HER and 6G TOPAN, stationed at JRR-3, Japan.
Co-aligned single crystals (64 pieces, 1.6~g mass in total) on the horizontal $[HK0]$ zone were used, and the total mosaicity was estimated to be $\sim$1.5~deg.
Both spectrometers adopt pyrolytic graphite crystals for the monochromator and analyser.
To evaluate the spin dynamics of FeGa$_2$S$_4$, we employed the fixed final energy $E_{\rm f} = 2.423$, 3.635, 6.7~meV for HER with cold neutrons, and $E_{\rm f} = 13.5$, 30.5~meV for TOPAN with thermal neutrons.
After optimising magnetic reflection and identifying the magnetic wavevector, temperature scans sitting on the peak top were performed.

%\section{Results \& Discussion}

\begin{figure}[t]
	\centering
	\includegraphics[width=\linewidth,bb=0 0 732 933]{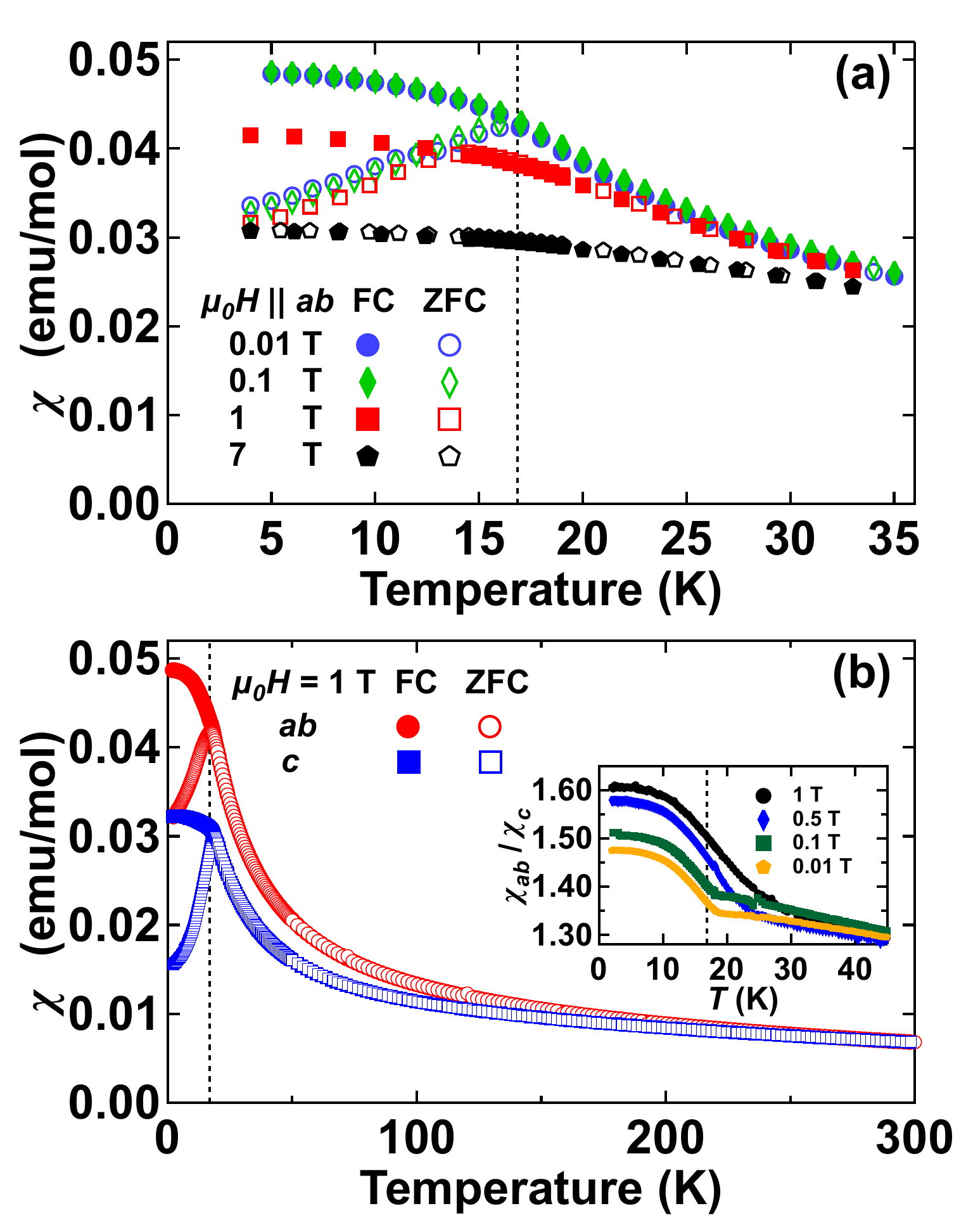}
	\caption{Temperature dependence of DC susceptibility (a) under various magnetic fields within the $ab$-plane at low-temperature regime, and (b) under $\mu_0H=0.1$~T within the $ab$-plane and along the $c$-axis at the whole temperature regime measured. The inset to (b) depicts the ratio $\chi_{ab}/\chi_c$ as a function of temperature. The dotted vertical lines indicate $T^{\ast}=16.87(7)$~K.}
	\label{DC}
\end{figure}

We first conducted DC magnetometry measurements on the single crystals to estimate the $T^{\ast}$ anomaly.
Measurements with various external magnetic fields ($\mu_0H$) parallel to the $ab$-plane were performed [Fig.~\ref{DC}(a)], and the $T^{\ast}$ anomaly is well characterised by the bifurcations between field-cooling (FC) and zero-field-cooling (ZFC) protocols.
$T^{\ast}$ was determined by a Gaussian peak fit to the obtained ZFC susceptibility.
The fit to the 0.01~T data yields $T^{\ast} = 16.87(7)$~K, which is in good accordance with the previously study~\cite{Nakatsuji2007PRL}.
Curie-Weiss fits to the data measured up to 300~K were performed on the 0.1~T data, giving $|\theta_{\rm W}|=155(3)$~K being close to the previously evaluated 160~K~\cite{Nakatsuji2007PRL}.
The obtained $T^{\ast}$ and $|\theta_{\rm W}|$ indicate relatively strong geometric frustration effects with the so-called frustration parameter of $f\equiv |\theta_{\rm W}|/T^{\ast}=9.2(2)$.
DC susceptibility with $\mu_0H\parallel ab$ and $\parallel c$ shows the difference in magnitude [Fig.~\ref{DC}(b)], implying the weak easy-plane anisotropy.
The magnetic anisotropy can roughly be evaluated as $\chi_{ab}/\chi_c$, which is up to 1.6 [inset to Fig.~\ref{DC}(b)], slightly higher than 1.3 from the previous study~\cite{Nakatsuji2007PRL}.

\begin{figure}[t]
	\centering
	\includegraphics[width=\linewidth,bb=0 0 869 1187]{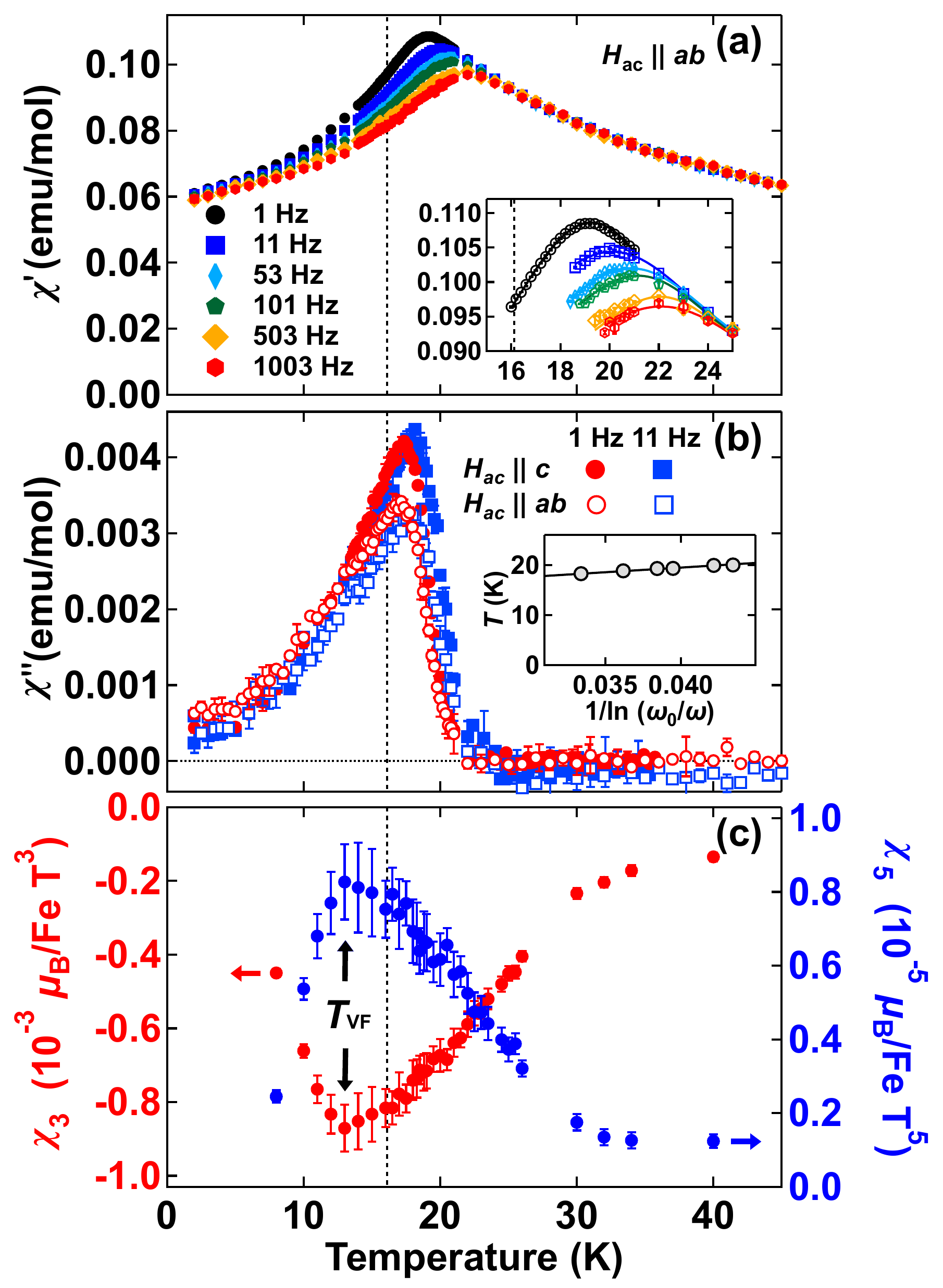}
	\caption{Temperature dependence of (a) the in-phase component of AC susceptibility within the $ab$-plane under $H_{\rm AC}=3$~Oe with various frequencies, (b) the out-of-phase component of AC susceptibility within the $ab$-plane and along the $c$-axis under $H_{\rm AC}=3$~Oe with 1 and 11~Hz, and (c) nonlinear susceptibilities. The insets to (a) and (b) depict Gaussian fits of the data and Vogel-Fulcher fit of the peak centre of AC susceptibility, respectively. The dotted vertical lines indicate $T^{\ast}=16.87(7)$~K.}
	\label{AC}
\end{figure}

Slow spin fluctuations across $T^{\ast}$ at the uniform $\vec{Q}\rightarrow 0$ limit with $\vec{Q}$ being the momentum transfer, were investigated by AC magnetometry measurements ranging from hertz to kilohertz.
Figure~\ref{AC}(a) displays the in-phase component of the AC susceptibility ($\chi^{\prime}$) with the AC magnetic field of $H_{\rm AC}=3$~Oe and without DC magnetic field.
The data clearly exhibit frequency-dependent peaks at around 20~K, accompanying cusps in the out-of-phase component ($\chi^{\prime\prime}$) [Fig.~\ref{AC}(b)].
Such behaviours, together with the FC/ZFC bifurcation, are reminiscent of a conventional SG transition~\cite{Mydosh1993CRC,Mezard1984PRL,Dotsenko1993PU}, however, this is not the case for FeGa$_2$S$_4$ as will be discussed later on.

To exemplify the possible glassiness of spins, we detailedly examine the frequency dependence.
The peak temperature, $T_{\rm low}(\omega)$ with $\omega$ being the angular frequency, is defined as the peak centre via the Gaussian peak fit [inset to Fig.~\ref{AC}(a)].
For the conventional SG transition, the relationship between $T_{\rm low}(\omega)$ and frequency, $f=\omega/(2\pi)$, can generally be described using the Vogel-Fulcher law~\cite{Vogel1921,Fulcher1925,Binder1986RMP}, 
\begin{equation}
	\omega=\omega_0\exp\left(-\frac{E_a}{k_{\rm B}(T_{\rm low}-T_{\rm VF})}\right),
	\label{VFfunc}
\end{equation}
where we adopt the constant $\omega_0 = 10^{13}$~Hz following the earlier studies~\cite{Tholence1980,Prejean1978,Dho2002}. 
The fit [inset to Fig.~\ref{AC}(b)] yields a Vogel-Fulcher temperature of $T_{\rm VF} = 12(3)$~K [Fig.~\ref{AC}(c)] and an activation energy of $E_a/k_{\rm B} = 187(11)$~K.
At $T=T_{\rm VF}$, spins are supposed to be frozen, where the spin relaxation time should diverge.
The evaluated $T_{\rm VF}<T^{\ast}$ suggests that spin fluctuations in FeGa$_2$S$_4$ do not immediately cease at $T^{\ast}$ but last to $\sim$12~K.
The Vogel-Fulcher parameter, $x = (T^{\ast} - T_{\rm VF})/T_{\rm VF} = 0.34$, stays within the typical range for conventional SG systems (0.03--1)~\cite{Tholence1984PBC,Anand2012PRB}, and the ratio $(T_{\rm VF} k_{\rm B})/E_a \ll 1$ aligns with the weak coupling regime~\cite{Shtrikman1981PRA}.

In further search for possible SG-like behaviour, we performed the magnetisation measurements to derive the third- ($\chi_3$) and fifth-order nonlinear susceptibilities ($\chi_5$).
The isothermal magnetisation was carefully measured at many temperatures to derive the nonlinear susceptibilities, and its field dependence was fit to the odd-order polynomial.
Figure~\ref{AC}(c) summarises the temperature dependencies of $\chi_3$ and $\chi_5$, and neither shows divergent behavior at $T^{\ast}$.
The nonlinear susceptibilities in the conventional SG systems should diverge at the bifurcation temperature~\cite{Suzuki1977PTP}, which starkly contrasts with our findings.
Instead, negatively and positively rounded peak formations can be found for $\chi_3$ and $\chi_5$, respectively. 
These are found at a temperature very close to $T_{\rm VF}$ involving a quasi-static spin state, which gives consistent views of the temporal correlation from the AC and nonlinear susceptibilities.
In addition to the absence of divergence in $\chi_3$ and $\chi_5$, the quadratic temperature dependence of the observed magnetic-specific heat at low temperatures~\cite{Nambu2008PRL} also contrasts with the conventional linear-$T$ behaviour known for SG systems.
These findings support that the $T^{\ast}$ anomaly in FeGa$_2$S$_4$ essentially differs from the conventional SG transition.

\begin{figure}[t]
	\centering
	\includegraphics[width=\linewidth,bb=0 0 1432 907]{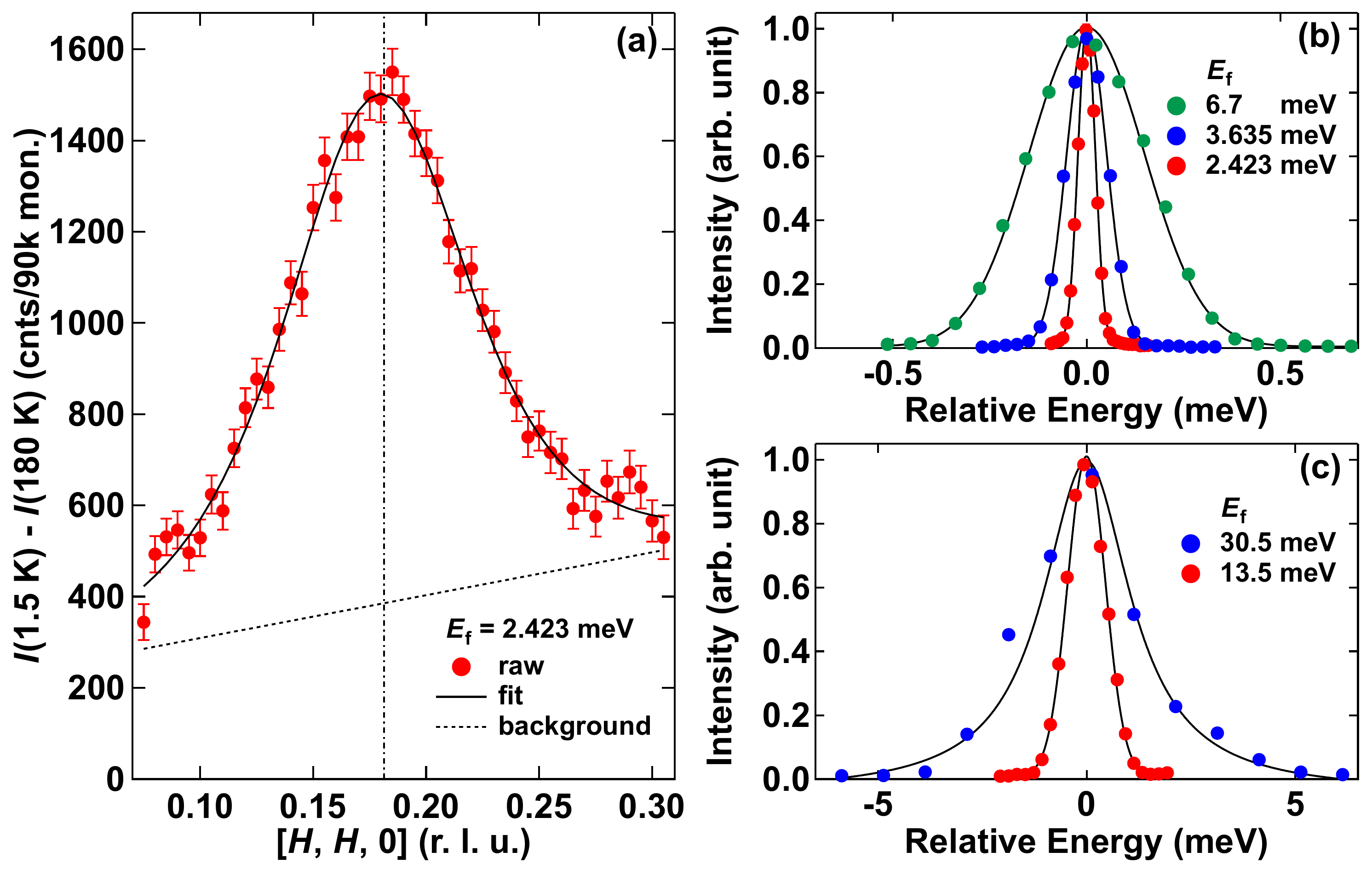}
	\caption{(a) The difference between elastic neutron scattering data taken at 1.5 and 180~K gives the magnetic signal. The dashed vertical and dotted lines indicate the peak centre at $\vec{q}_{\rm m}=(0.1814(7),0.1814(7),0)$~r.l.u. and the slope background, respectively. Energy resolution profiles were collected on (b) the cold-neutrons triple-axis spectrometer HER and (c) the thermal-neutrons triple-axis spectrometer TOPAN.}
	\label{Trip}
\end{figure}

To trace faster spin dynamics, neutron scattering experiments were carried out.
We first precisely evaluate the magnetic wavevector ($\vec{q}_{\rm m}$) using scans along the in-plane $[HH0]$ direction.
Figure~\ref{Trip}(a) shows the difference in elastic neutron scattering between 1.5 and 180~K collected on HER with $E_{\rm f}=2.423$~meV, which comes from purely magnetic signals.
We used a squared Lorentzian peak~\cite{Nakatsuji2005Sci,Stock2010} to fit the magnetic scattering,
\begin{eqnarray}
	I= A\frac{2\kappa^3/\pi}{\left[(\vec{Q}-\vec{q}_{\rm m})^2+\kappa^2\right]^2}+{\rm bkg.},
	\label{EqInc}
\end{eqnarray}
where $A$ is a coefficient, $\kappa\equiv 1/\xi$ denotes the inverse correlation length ($\xi$), and bkg. as the sloping background.

The fit using eq.~\ref{EqInc} results in an incommensurate magnetic wavevector $\vec{q}_{\rm m}=(0.1814(7),0.1814(7),0)$ r.l.u., corresponding to $65.3(2)^{\circ}$ rotated spin arrangement over neighbouring spins.
The observed magnetic scattering [Fig.~\ref{Trip}(a)] is actually not momentum resolution-limited, and estimated $\xi$ is about 3.2(7) times lattice spacing reflecting the short-ranged correlation in the compound.
The two-dimensional antiferromagnetic spin correlations revealed by $\vec{Q}$-dependent magnetic neutron scattering qualitatively differ from the conventional SG systems, where no $\vec{Q}$-dependence is anticipated~\cite{Mydosh1993CRC}.

The observed incommensurate $\vec{q}_{\rm m}$ can only be accounted for by the Hamiltonian with antiferromagnetic interactions up to the third-neighbours ($J_1$-$J_2$-$J_3$ [Fig.~\ref{Sam}(a)])~\cite{Guratinder2021PRB}.
On the contrary, magnetic interactions in NiGa$_2$S$_4$ are effectively modelled using ferromagnetic $J_1$ and antiferromagnetic $J_3$ interactions.
This leads to the wavevector $(0.155(3), 0.155(3), 0)$~r.l.u.~\cite{Stock2010} being smaller than the commensurate $(1/6, 1/6, 0)$~r.l.u. when dominated only by the antiferromagnetic $J_3$.
Determining the interaction strengths is beyond the scope of the present study and will be reported elsewhere.
We argue that only one $\vec{q}_{\rm m}$ is realised in our samples between 1.5 and 110~K, contrasting with the reported double-$\vec{q}_{\rm m}$ state below 5~K~\cite{Guratinder2021PRB}.
The separate behaviour may be owing to the absence of site disorder among cations in our samples, which requires further exploration.

\begin{figure*}[t]
	\centering
	\includegraphics[width=0.8\linewidth,bb=0 0 980 372]{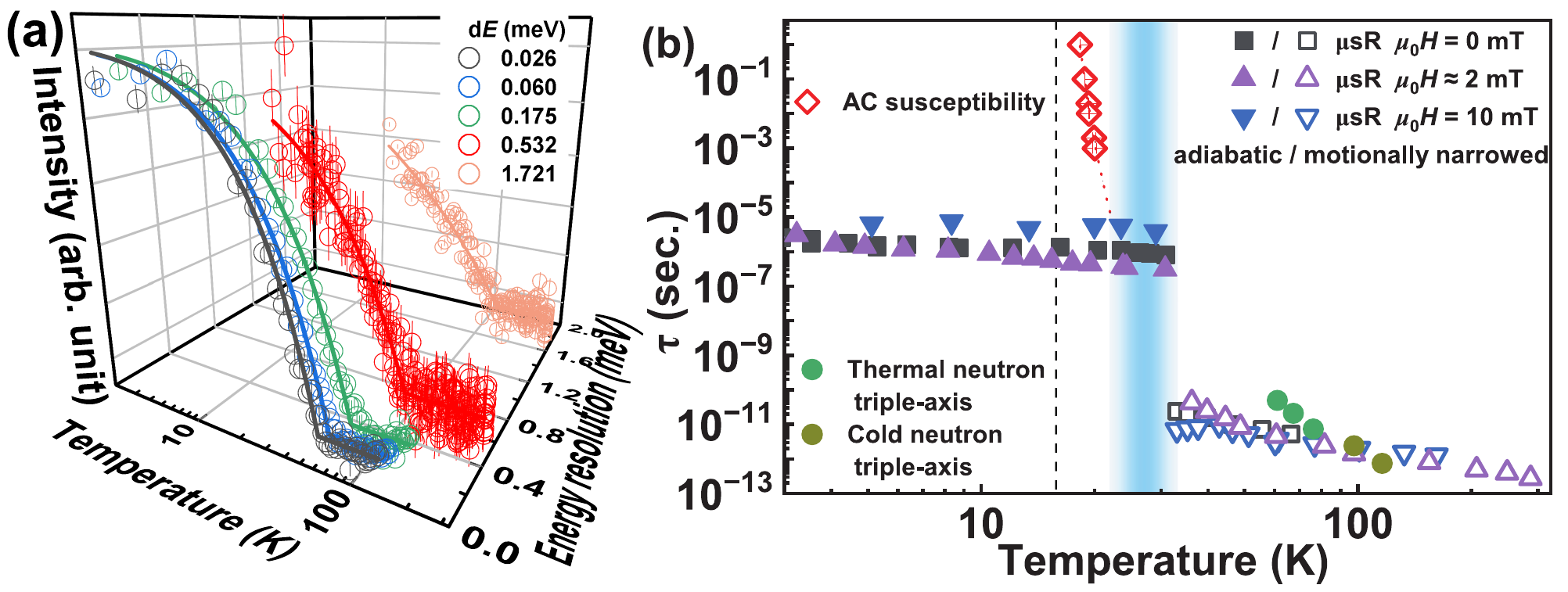}
	\caption{(a) Temperature dependence of neutron-scattering intensity in the elastic channel obtained on triple-axis spectrometers with separate $E_{\rm f}$ sitting on $\vec{q}_{\rm m}=(0.1814(7),0.1814(7),0)$~r.l.u. The solid curves are the fits (see main text) to the data. (b) Temperature dependence of characteristic spin relaxation time, $\tau$. The dotted red curve indicate the extrapolated Vogel-Fulcher fit to the higher temperature regime and the dashed black line indicates $T^{\ast}$. Data for the thermal- and cold-neutrons triple-axis spectrometers are determined as the onset temperature of (a). The $\mu$SR data are derived from the measured data in Refs.~\cite{Rotier2012PRB,Zhao2012PRB} for adiabatic ($T<T_{\mu}$) and montionally-narrowed ($T>T_{\mu}$) regimes. The shaded area indicates the slowly fluctuated regime for $22~\rm{K}<$ $T\leq T_{\mu}$.}
	\label{Dyn}
\end{figure*}

We then traced temperature variations of the magnetic scattering with changing spectrometer energy resolutions ($\delta E$).
The $\delta E$ of HER and TOPAN are determined by Gaussian fits to energy-scan data of incoherent scattering from the standard vanadium sample [Fig.~\ref{Trip}(b)] and FeGa$_2$S$_4$ single crystals [Fig.~\ref{Trip}(c)], respectively.
Figure~\ref{Dyn}(a) shows the temperature dependence of nominally elastic magnetic neutron scattering from FeGa$_2$S$_4$, where the intensity is measured sitting on $\vec{q}_{\rm m}$ for the temperature range of 2--170~K (HER) and 2--300~K (TOPAN).
The $\delta E$ sets the minimum spin relaxation time, $\tau=\hbar/\delta E$, via the uncertainty principle for spin fluctuations that contribute to intensity in the elastic channel.
The gradual appearance of elastic scattering below the onset temperature ($T_{\rm onset}$) thus denotes the development of spin correlations on a time scale beyond $\tau=\hbar/\delta E$.

The temperature dependence of the elastic magnetic scattering is fit using a phenomenological formula,
\begin{equation}
	I=\cases{B\left(1-\left(\frac{T}{T_{\rm onset}}\right)^{\alpha}\right)^{\beta}+{\rm Const.}&for $T<T_{\rm onset}$,\\
	{\rm Const.}&for $T>T_{\rm onset}$,\\}
	\label{Phe}
\end{equation}
with $B$ as a coefficient and appropriate exponents of $\alpha$ and $\beta$.
Figure~\ref{Dyn}(b) summarises $\tau$ of FeGa$_2$S$_4$ as a function of temperature determined by several experimental techniques.
The downward shift in $T_{\rm onset}$ with tighter energy resolution indicates that the time scale for spin fluctuations increases with decreasing temperature.

In Fig.~\ref{Dyn}(b), the $\mu$SR data are obtained from the previously measured data~\cite{Rotier2012PRB,Zhao2012PRB}.
The $\tau$ is derived in the cases for slow (adiabatic~\cite{Hayano1979}) at $T<T_{\mu}$ and rapid (motionally-narrowed~\cite{Maclaughlin2008PRB,Nambu2015PRL}) fluctuations at $T>T_{\mu}$,
\begin{eqnarray}
	\lambda_s=\cases{\frac{2}{3}\tau^{-1}&for $T<T_{\mu}$,\\
	(\gamma_{\mu} B_{\rm{loc}})^2\tau&for $T>T_{\mu}$,\\}
\end{eqnarray}
where $\gamma_{\mu}=2\pi \times 135.5$~MHz/T is the muon gyromagnetic ratio, and the local field $B_{\rm loc} = 0.2$~T is used~\cite{Reotier2009JPCS}.
Reflecting techniques probing separate aspects of spin correlations, $\tau$ from the $\mu$SR data are considerably shorter than the data inferred from neutron scattering at $T<5T^{\ast}$.
This highlights the difference in detecting correlations at a characteristic $\vec{Q}$ by neutrons and for $\vec{Q}$-averaged by muons.
The difference at low temperatures is prominent, where the longer relaxation times for $\vec{Q}\rightarrow 0$ AC susceptibility are realised compared to local probes of $\mu$SR.

Figure~\ref{Dyn}(b) underlines a slowly fluctuated regime with spin fluctuation of $\sim 10^{-5}$~sec (shaded area).
This sort of spin-liquid-like state takes place below $T_{\mu}$ upon cooling, and it lasts at least down to approximately 22~K, where the extrapolated Vogel-Fulcher curve merges into the $\mu$SR data.
Whereas such a behaviour resembles the result of NiGa$_2$S$_4$~\cite{Nambu2015PRL}, the temperature regimes of the two compounds interestingly deviate from each other.
The slowly fluctuated regime lies below the FC/ZFC bifurcation temperature (again defined as $T^{\ast}$) for NiGa$_2$S$_4$ ($4~\mathrm{K}<T\leq T^{\ast}$)~\cite{Nambu2015PRL}, on the other hand, it is above for FeGa$_2$S$_4$ ($22~\mathrm{K}<T\leq T_{\mu}$).
Furthermore, if the scaling over $T/T_{\mu}$ was applied on the horizontal axis of Fig.~\ref{Dyn}(b), the regime lasts down to $0.71T_{\mu}$ for FeGa$_2$S$_4$ and $0.44T_{\mu}$ for NiGa$_2$S$_4$, highlighting more stabilised fluctuations by quantum effects.
The slow spin fluctuation is also observed in the triangular antiferromagnet HCrO$_2$~\cite{Somesh2021PRB}.
It shows the magnetic long-range order at $T_{\rm N}$, but the interlayer frustration interestingly stabilises the slow fluctuation even below $T_{\rm N}$.
Information on temporal spin correlations is becoming increasingly important for understanding frustrated magnetism.

Further differences can be found in the two compounds for temperature-dependent behaviours of $\chi^{\prime}$, $\chi_3$ and $\chi_5$~\cite{Nambu2015PRL}.
These might relate to the separate spin sizes and the remaining degree of freedom in the $t_{2g}$ orbital in FeGa$_2$S$_4$.
Despite their structural simplicity and considerable theoretical efforts, no theory has hitherto fully accounted for the unusual spin dynamics and such differences, however some of them succeed in phenomenologically explaining our findings.

One is the topological binding-unbinding crossover of vortices formed by vector spin chirality~\cite{Kawamura1984JPSJ}.
The theory computes the crossover temperature as $T_{\rm v}=0.285|J|S^2\sim 14.7$~K close to $T^{\ast}$ of FeGa$_2$S$_4$, and predictions reproduce the observed consecutive anomalies in the susceptibility and specific heat.
Another is the spin nematic state~\cite{Lauchli2006PRL,Tsunetsugu2006JPSJ}, which conveys an order of magnetic quadrupole moments.
This scenario can explain the macroscopic quantum effects at low temperatures in $A$Ga$_2$S$_4$~\cite{Nambu2008PRL,Nambu2011,Nambu2006JPSJ}.
Both theories arise from the higher-order degree of freedom compared to the dipole moments, but they do not take into account interactions beyond the nearest-neighbour interactions. 
Including further neighbour interactions, quantum effects, and orbital degrees in such theoretical models would fully explain our findings, including the unusual temporal spin correlations and the resultant incommensurability.

%\section{Conclusions}

To summarise, we have demonstrated the temporal spin correlations in FeGa$_2$S$_4$ over 13 decades.
Despite some similarities to conventional SG systems, the magnetometry measurements clearly depict that the glass transition is not the case for the $T^{\ast}$ anomaly.
Neutron and muon results instead identify the slowly fluctuated spin state near $T^{\ast}$, a rare realisation of the spin-liquid state with quantitative information.
Low-frequency and low-temperature spin fluctuations in FeGa$_2$S$_4$ and NiGa$_2$S$_4$ are uncommon in geometrically frustrated magnets.
The present study will spur further theoretical work in developing models of frustrated magnets that can describe such temporal correlations.

We thank K. Aoyama, C. Broholm, K. Guratinder, M. Imada, H. Kawamura, S. Nakatsuji, C. Stock, O. Tchernyshyov, and H. Tsunetsugu for their valuable discussions, and J. Okada for his assistance during the SEM-EDX measurements.
This work was supported by the JSPS (Nos.~21H03732, 22H05145, 17H06137), FOREST (No.~JPMJFR202V) and SPRING from JST, and the Graduate Program in Spintronics at Tohoku University.

\end{document}